\documentclass[a4paper,10pt]{article}
\usepackage[utf8]{inputenc}
\usepackage{amssymb,amsmath,amsfonts}
\usepackage{graphicx}
\usepackage{epstopdf}
\usepackage{bm}
\usepackage{latexsym}
\usepackage{graphics, setspace}

\newcommand{\mathsym}[1]{{}}
\newcommand{\unicode}[1]{{}}

\usepackage{graphicx}
\usepackage{bm}
\usepackage{epstopdf}
\usepackage[left=2cm,right=2cm,top=2cm,bottom=2.5cm]{geometry}

\newcommand{\be}{\begin{equation}}
\newcommand{\ee}{\end{equation}}
\newcommand{\bea}{\begin{eqnarray}}
\newcommand{\eea}{\end{eqnarray}}
\newcommand{\<}{\langle}
\renewcommand{\>}{\rangle}

\def\<{\langle}
\def\>{\rangle}
\def\(({\left(}
\def\)){\right)}

\usepackage{subfig}
\usepackage{color}

\title{One-loop topological expansion for spin glasses in the large connectivity limit}
\author{Maria Chiara Angelini$^{1}$ \and Giorgio Parisi $^{1,2,3}$ \and Federico Ricci-Tersenghi$^{1,2,3} $  \\                 
  $^{1}$ Dipartimento di Fisica, Sapienza Universit\`a di Roma, Piazzale A. Moro 2, I-00185, Rome, Italy\\
  $^{2}$ Nanotec-CNR, UOS Rome, Sapienza Universit\`a di Roma, Piazzale A. Moro 2, I-00185, Rome, Italy,\\
 $^{3}$ INFN-Sezione di Roma 1, Piazzale A. Moro 2, 00185, Rome
}
\date{}
\begin{document}
\maketitle

\abstract{We apply for the first time a new one-loop topological expansion around the Bethe solution to the spin-glass model with field in the high connectivity limit,
following the methodological scheme proposed in a recent work. 
The results are completely equivalent to the well known ones, found by standard field theoretical expansion around the fully connected model (Bray and Roberts 1980,
and following works). 
However this method has the advantage that the starting point is the original Hamiltonian of the model, with no need to define an associated field theory, nor to
know the initial values of the couplings, and the computations have a clear and simple physical meaning. 
Moreover this new method can also be applied in the case of zero temperature, when the Bethe model has a transition in field, contrary to the fully connected model that is 
always in the spin glass phase. Sharing with finite dimensional model the finite connectivity properties, the Bethe lattice is clearly a better starting point for an expansion
with respect to the fully connected model. The present work is a first step towards the generalization of this new expansion 
to more difficult and interesting cases as the zero-temperature limit, where the expansion could lead to different results with respect to the standard one.}

\section{Introduction}
Spin Glasses (SGs) are models whose mean field (MF) version \cite{SK} undergoes a phase transition, crossing 
a critical line in the temperature-field $(T-h)$ plane. 
The solution of the MF problem sees the introduction of replicas of the original system as a mathematical trick to perform computations. The resulting Hamiltonian is symmetric under
replica exchanges. However, quite surprising, one finds that in the low-temperature spin-glass phase the replica symmetry is broken.
While the MF behavior of the model is completely under control \cite{FRSB}, also from a rigorous viewpoint \cite{PAN}, we still do not have a confirmed theory for the finite dimensional version.
In particular, there is not agreement both on the upper and lower critical dimension, looking at theoretical, numerical and experimental data 
\cite{Moore2011,ReplyMoore2011,SG_MK,Yaida,LR1,LR2,LR3,Janus,experimental}.

The project to perform a renormalization group (RG) analysis is an old one.
The spin glass transition in zero-field was already studied within the RG by Harris et al. \cite{Harris}, and in field by Bray and Roberts \cite{BR}, limiting at the
sector associated with the critical eigenvalue, the so called \textit{replicon}.
Their one-loop analysis was then repeated adding the other sectors, longitudinal and anomalous ones, in refs. \cite{Temesvari1,Temesvari2,Temesvari3,Temesvari4}.
In a recent work also the two loop computation in a field has been performed \cite{Yaida}, suggesting the possibility of a non-perturbative fixed point.

These works are expansions around the Fully Connected (FC) mean field model. 
They start studying  the symmetric phase,  approaching  the transition from the high-temperature side, where replica symmetry holds.
Thus the replica symmetric Lagrangian is written, that in its most complete version has three bare masses and eight cubic couplings involving the
replica fields, which correspond to all the possible invariants under the replica symmetry.
At this point one can perform a renormalization \'a la Wilson, integrating the degrees of freedom over an infinitesimal momentum shell, extracting the leading, one-loop, order approximation 
in $\epsilon=6-d$. 
Although the scheme is clear, the computation is highly technical also for the algebraic viewpoint. 

Recently a new loop expansion around the mean field Bethe solution was proposed in ref. \cite{Mlayer}. The new method can be applied to each model that is well defined on a Bethe lattice.
In this paper, we apply for the first time this new expansion to the SG in a field. 
We restrict ourself to the limit of high connectivity $z\rightarrow\infty$ to perform computations analytically. 
We compute the 1th order correction, and we show that in the $T>0$ region this new expansion is completely equivalent to the field theoretical one, 
recovering the results of Bray and Roberts \cite{BR}.
However it has the advantage that the starting point is the original Hamiltonian of the model, with no need to define an associated field theory, nor to
know the initial values of the couplings, and the computations have a clear and simple
physical meaning: while in standard field theory Feynmann diagrams have no special meaning, here the important diagrams have a geometrical interpretation. 
Moreover, the expansion is around Bethe lattice that has finite connectivity, an important characteristic shared with finite dimensional systems.
Even if in this work we obtain the same results as standard RG around the fully-connected model, 
we will discuss the differences that could arise in the two methods in particular situations.
This work is first of all a verification of the correctness of the method proposed in ref. \cite{Mlayer} and it is a first step towards the generalization 
of this new expansion to more complicated cases: finite small connectivity and zero temperature.

\section{Expanding around the Bethe lattice solution}

In this section we just recap the results of ref. \cite{Mlayer}, while in following sections we will apply for the first time these results to the SG model in a field.
Starting from a $D$ dimensional system, the $M$-layer construction of ref. \cite{Mlayer} consists of taking $M$ copies of the original model and rewire them. 
In the $M\rightarrow\infty$ limit, the rewiring procedure leads to the Bethe lattice. One can then expand the observables in powers of $\frac{1}{M}$.
In particular, we will focus our attention on the correlation functions, connected over the disorder, let's name them $G$. 
The $\frac{1}{M}$ expansion results in a topological expansion in the number of loops.
At order $\frac{1}{M}$, the correlation function between the origin and a point $d$ in the $D$ dimensional model results to be:
$G(d)=\sum_{L=1}^{\infty}{\cal B}(d,L) g^B(L)$, where ${\cal B}(d,L)$ is the number of non-backtracking walks that go from the origin to the point $d$ in the original $D$ dimensional model
and $g^B(L)$ is the correlation between points at distance $L$ on a Bethe lattice.

\begin{figure}
\centering
\includegraphics[width=0.9\columnwidth]{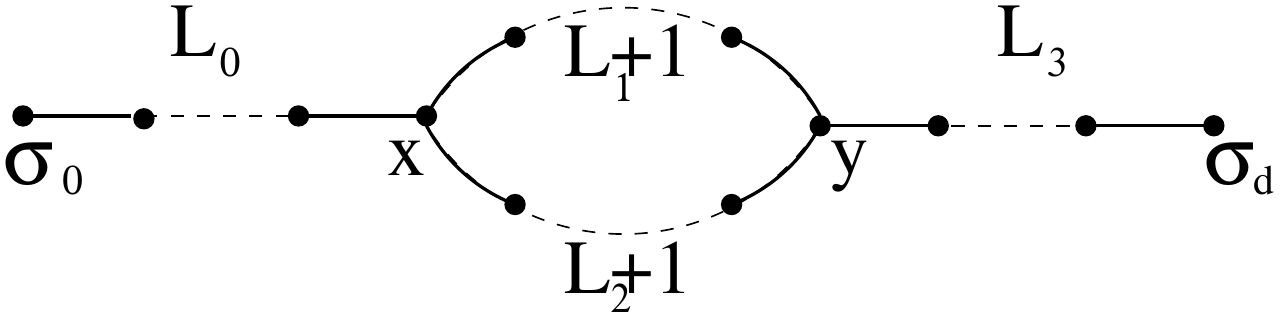}
\caption{\label{Fig:loop} Spatial loop $\mathcal{L}$ that gives the first correction to the bare correlation functions in the expansion around the Bethe solution.}
\end{figure}
At order $\frac{1}{M^2}$, the correlation function on the original system receive a leading contribution that is 
the product of the so-called \textit{line-connected} observable $g^{lc}(\cal L)$ computed on a Bethe lattice, 
in which it has been manually injected a loop ${\cal L}$ of the type in Fig. \ref{Fig:loop},
multiplied by the number of such a structure ${\cal L}$ present on the original model.
The line-connected observable is just the observable computed on a Bethe lattice with the loop minus the observable computed on the two paths $L_0,L_1+1,L_3$ and $L_0,L_2+1,L_3$
considered as independent.

The quantities $g^B(L)$ and $g^{lc}(\cal L)$ are model dependent.
In the following we will compute them for the Spin Glass in a field.
To make the computation analytically feasible, we will compute things on a Bethe lattice in the high connectivity limit, at temperature $T>0$.
However things can be computed in finite connectivity and even at $T=0$ using numerically the Belief Propagation equations. This will be the subject of a subsequent paper.

We just want to recall that one could perform the same $M$-layer construction around the fully connected model instead of the Bethe lattice. In the former approach, the leading
divergences at each order are exactly given by the corresponding terms in the loop expansion of the continuum  field theory.
In ref. \cite{Mlayer} it is claimed that, if the critical behaviour of the Fully connected model and of the Bethe lattice model is the same, than also the two expansions will lead to the
same results. We will see that this is exactly the case for the SG with field at $T>0$.

\section{Model and definitions}

To be concrete, we are interested in the SG model in field, that has the following Hamiltonian:
$$
H(\{{\sigma\}})=-\sum_{i=1,N} \sigma_i h_i^R -\sum_{ij\in E} \sigma_i \sigma_j J_{ij}\,,
$$
where $E$ is the set of the edges of the lattice. On this model we should compute the quantities $g^B(L)$ and $g^{lc}(\cal L)$ introduced in the previous section.

We will consider the model on a Bethe lattice (for definiteness on a random $z$-regular lattice) in the large connectivity limit, i.e.\ $z$ large: 
we will keep the leading terms and we will neglect the $1/z$ corrections. Only at the end we will perform the limit $z\to\infty$.  
In this limit, computations are easier and the model has the same properties of the Sherrington Kirkpatrick one \cite{SK}. 
The procedure of first computing the results for finite $z$ in the thermodynamic limit ($N\to\infty$) and later send $z$ to infinity makes the physical approach much clearer. 

The couplings are i.i.d.\ random variables extracted from a distribution with the following properties: $\overline{J_{ij}}=0$, $\overline{J_{ij}^2}=1/z$.
Higher order moments are irrelevant in the $z\to\infty$ limit. We have indicated by $h_i^R$ the field on the site $i$. 
It can be either a local random field extracted from a given distribution or a spatially uniform field. The physics is
equivalent in the two cases. 
For simplicity here we consider the case where the fields $h_i^R$  are Gaussian variables with zero average and finite variance $v_h$ 
\footnote{The attentive reader could notice that the variance $v_h$ never appears in the following. This is not because things do not depend on $v_h$, but because the dependence is 
hidden in the definitions of the magnetizations: $m^2\equiv \overline{\langle \sigma\rangle^2}$ and higher moments will implicitly depend on $v_h$.}.
With standard notation, we indicate with $\langle\cdot\rangle$ the thermal average and with $\overline{\:\cdot\:}$ the average over the disorder (random couplings and fields).

In the thermodynamic limit we will compute different kinds of correlation functions first between point at distance $L$ on a standard Bethe lattice; 
this will lead to the \textit{bare} propagator
and we will compute its exact expression with two different methods in the high-temperature region: the replica method and the cavity method.
Then we will compute the first correction to this result due to the presence of one spatial loop.
The limit $z\rightarrow\infty$ allows us to compute all the quantities analytically. For finite connectivity, one could 
compute everything numerically using Belief-Propagation as usually done on Bethe lattices.

\section{The replica computation of line-correlations}\label{Sec:line_rep}

In the high connectivity limit (that corresponds to small couplings limit) we can expand the replicated partition function as:
\begin{align}
\nonumber
\overline{Z^n}=&\overline{\sum_{\{{\sigma\}}}e^{\beta\sum_a\sum_i\sigma_i^ah_i^R}e^{\beta\sum_a\sum_{ij}\sigma_i^a\sigma_j^aJ_{ij}}}=\\
\simeq&\overline{\sum_{\{{\sigma\}}}e^{\beta\sum_a\sum_i\sigma_i^ah_i^R}\prod_{ij}\((1+\beta\sum_a\sigma_i^a\sigma_j^aJ_{ij}+\frac{\beta^2}{2}\sum_{a,b}\sigma_i^a\sigma_i^b\sigma_j^a\sigma_j^bJ^2_{ij}\))}
\label{eq:Zn_largez}
\end{align}

The neglected terms give sub-leading contributions for large $z$.

As usual in SG computations, $a,b,c,\ldots\in[1,n]$ indicate the replica index, where the replicas are independent copies of the system with the same disorder realization.

A Bethe random regular graphs in the $N\to\infty$ limit becomes locally loop-less. The distance on the graph between two generic points (i.e. the length of the shortest path between them) 
is of order $\log(N)$. We are interested in the computation of the correlation functions of spins that are on points at a distance $L$ between them, 
in the limit where $N$ goes to infinity at fixed $L$. 
In this limit with probability one there is an unique path (of finite length) connecting them so the computation can be done on a single line. 

In general we will be interested in correlation functions that are connected
with respect to the disorder. At this end it is convenient to compute 
\begin{equation}
 G_{a,b;c,d}(L)\equiv\overline{\langle\sigma_0^a\sigma_0^b\sigma_L^c\sigma_L^d\rangle}-\overline{\langle\sigma_0\rangle^2}\cdot \overline{\langle\sigma_L\rangle^2}\,,
\end{equation}
from which we can extract connected and disconnected (with respect to thermal average) correlation functions.
In the following, for simplicity of notation, we will indicate with $\overline{ (\cdot) }^c$ the correlations connected with respect to the disorder: 
$\overline{\langle\sigma_0^a\sigma_0^b\sigma_d^c\sigma_d^d\rangle}^c\equiv 
\overline{\langle\sigma_0^a\sigma_0^b\sigma_d^c\sigma_d^d\rangle}-\overline{\langle\sigma_0\rangle^2}\cdot\overline{\langle\sigma_d\rangle^2}$.
We define the matrix $T\in \mathbb{R}^{n(n-1)\times n(n-1)}$ such as $T_{ab,cd}(i)\equiv\frac{\beta^2}{2}\overline{\langle\sigma_i^a\sigma_i^b\sigma_i^c\sigma_i^d\rangle}$.
We define $T$ only for $a\neq b$, $c\neq d$, following what is usually done for the matrix $Q_{ab}(i)=\langle\sigma_i^a\sigma_i^b\rangle$ in replica calculations
\footnote{Please be careful to not confuse $T$ with a tensor. We could use two superindices $i,j\in[1,n(n-1)]$ instead of the couples $ab$, $cd$. However we choose this notation
because it will be useful to define the different types of correlation functions in the following.}.
Remembering that $\overline{J_{ij}}=0$, $\overline{J_{ij}^2}=\frac{1}{z}$,  
from eq. (\ref{eq:Zn_largez}) we find that:
\begin{equation}
G_{a,b;c,d}(L)=\frac{2}{\beta^2z^L}\left[\prod_{i=0}^LT(i)\right]_{ab,cd}.
\label{eq:corr_d}
\end{equation}
Please notice that all the $i\in[0,L]$ are present in eq. (\ref{eq:corr_d}). In fact on a Bethe lattice, two spins are linked just by a path. If the link between two spins is cut, 
the spins become disconnected. This means that if a coupling is 0, all the correlation functions between the two spins linked by that coupling are zero. This automatically implies that
a correlation function should be proportional to the product of all the couplings on the path between the two spins.

From eq. (\ref{eq:corr_d}), we need to compute powers of the matrix $T$.
It is easy to show that:
\small
\begin{align}
\label{eq:T1}
T_{ab,cd}(i)=&\frac{\beta^2}{2}\overline{\langle\sigma_i^a\sigma_i^b\sigma_i^c\sigma_i^d\rangle}=\\
\nonumber
=&\frac{\beta^2}{2}\cdot 
\begin{cases} 
 1    &\text{ if }a=c\text{ , }b=d\text{ or if }a=d\text{ , }b=c \\
 m_2 &\text{ if }a=c\text{ or }b=d\text{ or }a=d\text{ or } b=c \\
 m_4 &\text{ if } a\neq b\neq c\neq d
 \end{cases}
 \end{align}
\normalsize
with $m_2=\overline{\langle\sigma_i\rangle^2}$ and $m_4=\overline{\langle\sigma_i\rangle^4}$.
Let us just mention that for eq. (\ref{eq:corr_d}) to hold, $T$ should be defined as the so-called ``cavity'' average:
the average over the rest of the system with the exception of the neighbouring spins on the considered line. However, in the large $z$ limit, cavity averages are equal to standard averages (see
also the Supplementary Material).

Eq. (\ref{eq:T1}) can be written in the form:
\begin{align} 
 \nonumber
 T_{ab,cd}=&\frac{\beta^2}{2}\cdot\left[m_4+(m_2-m_4)(\delta_{ad}+\delta_{bc}+\delta_{bd}+\delta_{ac})+\right.\\
 &\left.+(1-2m_2+m_4)(\delta_{ac}\delta_{bd}+\delta_{ad}\delta_{cb})\right].
\label{Eq:T}
\end{align}


In order to compute correlation functions, we need to compute  powers of $T$; For this reason, we proceed to the diagonalization of $T$. 
The whole calculation is explained in the Supplementary Material. Here we just sketch the main steps and state the final result.
First of all we look for eigenvalues and eigenvectors of $T_{ab,cd}$, of the form:
$\sum_{cd}T_{ab,cd}\psi_{cd}=\lambda\psi_{ab}$.
Because of the symmetry of the matrix $T$ under permutations of the replica indices, we know that there are three symmetry classes of eigenvectors 
(in an analogous way to what one does when looking to the stability of the Sherrington Kirkpatrick solution for the FC model \cite{dAT}) and three associated eigenvalues:

In the limit $n\rightarrow 0$ the first two eigenvalues (longitudinal and anomalous) are $\lambda_{L/A}=\beta^2(3m_4-4m_2+1),$
while the third one (replicon) is $\lambda_R=\beta^2(1-2m_2+m_4).$

We just want to point out that the eigenvalues are not the same ones as the usual ``replicon, anomalous, longitudinal'' eigenvalues 
that comes out from the diagonalization of the Hessian in ref. \cite{dAT}, but we called them in the same way because they identify the same sub-spaces with the same replica symmetries.
In particular at the spin-glass transition the usual replicon goes to zero, while $\lambda_R$ as defined in this paper goes to $\lambda_R=1$ leading to the 
divergence of the spin-glass susceptibility (see the following Section).

At this point, we construct the projectors on the sub-spaces of the eigenvectors and write $T$ as a combination of the projectors. In this representation, it is 
easy to compute powers of $T$. A special care should be taken in performing the limit $n\rightarrow0$, because of the degeneration of $\lambda_L$ and $\lambda_A$. 
The final result is:
\small
\begin{align*}
T^L(n=0)=&\frac{\beta^2}{2}L\lambda_{L/A}^{L-1}\((3m_4-2m_2\))R+\\
&+\lambda_{L/A}^L\((-\frac{R}{2}-Q\))+\lambda_R^L\((\frac{R}{2}+Q+P\)).
\end{align*}
\normalsize
where we defined the matrices $R_{ab,cd}=1$, $Q_{ab,cd}=\frac{1}{4}\left[\delta_{ac}+\delta_{ad}+\delta_{bc}+\delta_{bd}\right]$, 
$P_{ab,cd}=\frac{1}{2}\left[\delta_{ac}\cdot\delta_{bd}+\delta_{bc}\cdot\delta_{ad}\right]$.

For a spin-glass model, for each realization of the the system different correlation functions can be defined. 
Because of the symmetry of the coupling distribution, the average over the realizations of all the ``linear'' correlations will be zero, and the relevant ones
will be the squared correlations. We will define three main correlations averaged over the thermal noise:
\begin{itemize}
 \item The total correlation: $\langle\sigma_0 \sigma_L\rangle$
 \item The disconnected correlation: $\langle\sigma_0\rangle\langle\sigma_L\rangle$
 \item The connected correlation: $\langle\sigma_0\sigma_L\rangle_c=\langle\sigma_0\sigma_L\rangle-\langle\sigma_0\rangle\langle\sigma_L\rangle$
\end{itemize}
Obviously only two of these correlations are linearly independent.

With these two-spins correlations, we can build different squared correlations:
\begin{itemize}
 \item  The total-total correlation at distance $L$:

\begin{align}
\nonumber
\overline{\langle\sigma_0\sigma_{L}\rangle^2}^c=&\frac{2}{\beta^2z^{L}}\lim_{n\rightarrow 0}\left[\frac{1}{n(n-1)}\sum_{a\neq b}\((T^{L+1}\))_{ab,ab}\right]=\\
=&\frac{2}{\beta^2z^{L}}\((\frac{3}{2}\lambda_R^{L+1}-\lambda_{L/A}^{L+1}+(L+1)\lambda_{L/A}^{L}\frac{\beta^2}{2}(3m_4-2m_2)\)).
\label{Eq:corrTT_rep}
\end{align}
\item  The disconnected-disconnected correlation at distance $L$:
\begin{align}
 \nonumber
\overline{\langle\sigma_0\rangle^2\langle\sigma_{L}\rangle^2}^c=&\frac{2}{\beta^2z^{L}}\lim_{n\rightarrow 0}\left[\frac{1}{n(n-1)(n-2)(n-3)}\sum_{a\neq b \neq c\neq d}\((T^{L+1}\))_{ab,cd}\right]=\\
=&\frac{2}{\beta^2z^{L}}\((\frac{\lambda_R^{L+1}}{2}-\frac{\lambda_{L/A}^{L+1}}{2}+(L+1)\lambda_{L/A}^{L}\frac{\beta^2}{2}(3m_4-2m_2)\)).
\label{Eq:corrDD_rep}
\end{align}
\item The total-disconnected correlation at distance $L$:
\begin{align}
 \nonumber
\overline{\langle\sigma_0\sigma_{L}\rangle\langle\sigma_0\rangle\langle\sigma_{L}\rangle}^c=&\frac{2}{\beta^2z^{L}}\lim_{n\rightarrow 0}\left[\frac{1}{n(n-1)(n-2)}\sum_{a\neq b\neq c}\((T^{L+1}\))_{ab,ac}\right]=\\
=&\frac{2}{\beta^2z^{L}}\((\frac{3 \lambda_R^{L+1}}{4}-\frac{3\lambda_{L/A}^{L+1}}{4}+(L+1)\lambda_{L/A}^{L}\frac{\beta^2}{2}(3m_4-2m_2)\)).
\label{Eq:corrTD_rep}
\end{align}
\item The connected-connected correlation at distance $L$, whose expression can be obtained as a combination of the previous ones:

\begin{align}
 \nonumber
\overline{\langle\sigma_0\sigma_{L}\rangle_c^2}=&\overline{\((\langle\sigma_0\sigma_{L}\rangle-\langle\sigma_0\rangle\langle\sigma_{L}\rangle\))^2}=\overline{\langle\sigma_0\sigma_{L}\rangle^2}+\\
&\nonumber -2\overline{\langle\sigma_0\sigma_{L}\rangle\langle\sigma_0\rangle\langle\sigma_{L}\rangle}+\overline{\langle\sigma_0\rangle^2\langle\sigma_{L}\rangle^2}=\\
=&\frac{1}{\beta^2z^{L}}\lambda_R^{L+1}.
\label{Eq:corrCC_rep}
\end{align}

\end{itemize}
Also in this case only three of these correlations are linearly independent.
\footnote{The expression for the connected correlation in the Bethe lattice is obtained in this paper from a large $z$ expansion. 
However we numerically checked that also for small $z$ eq. (\ref{Eq:corrCC_rep}) is valid substituting $z$ with $z-1$ (that is equivalent in the large $z$ limit). }

Others correlations can be readily obtained from the previous one by integration by part. 
In the Supplementary Material, we show how to obtain the connected and disconnected bare correlation functions in a cavity approach leading to the same results.

\section{The dominant contribution in the correlation functions} \label{Sec:susc}

To build the susceptibility associated with a given correlation function $C(0,L)$ in a Bethe lattice (where $C$ can be one among the correlation functions considered in the previous section), 
one should sum over all the spins that are at distance $L$ from the spin 0, and then over all the distances $L$:
\be
\chi_C^B\propto\sum_{L=1}^{\infty}\mathcal{N}_L C(0,L),
\label{eq:chi}
\ee
where $\mathcal{N}_L=z(z-1)^{L-1}$ is the number of spins at distance $L$ from a given spin in a Bethe lattice with connectivity $z$. 
The SG transition line is commonly associated to the divergence of the SG susceptibility, that is the susceptibility associated to the connected correlation function.
Substituting eq. (\ref{Eq:corrCC_rep}) and the expression for ${\cal N}_L$ in eq. (\ref{eq:chi}), we discover that the critical line is identified by $\lambda_R=1$
\footnote{Indeed one can check that in the limit of $z\rightarrow\infty$ $\lambda_R$, as defined in this paper, is deeply related to the replicon eigenvalue $\lambda$
of eq. (15) from ref. \cite{dAT}. Using eq. (11) of ref. \cite{dAT}, it can be demonstrated that the relation $\big(\frac{KT}{J}\big)^2 \lambda=-\lambda_R+1$ holds.
The usual SG line associated with $\lambda=0$ translates in $\lambda_R=1$}.
Looking at the eigenvalues, we can numerically check that $\lambda_{L/A}<\lambda_R$. 
All the bare correlation functions, as shown in the two precedent sections, have a term proportional to $\lambda_R^L$, that is thus the dominant one.
Recovering the result of standard theory, the critical behavior of all the correlation functions is the same because all depend on the only critical eigenvalue \cite{DedominicisGiardina}.
The susceptibility associated to the different correlations, computed at the critical point, is divergent at the critical line. 
On the Bethe lattice, this divergence is a consequence of the exponential decay of the correlations multiplied by the exponential numbers of neighbours at a given distance.
Until now we have computed $\chi_C^B$ on the Bethe lattice.
If now we want to use the Bethe approximation to compute the susceptibility in a $D$ dimensional lattice,
following ref. \cite{Mlayer} (recall Sec. ``Expanding around the Bethe lattice solution''), we should replace $\mathcal{N}_L$ in eq. (\ref{eq:chi}) with the total
number ${\cal B}(d,L)$ of non-backtracking paths of length $L$ between two points at distance $d$ on the original lattice, and add a sum over $d$.
For large $d$ and $L$ 
\begin{equation}
{\cal B}(d,L) \propto (2D-1)^L |L|^{-D/2}\exp(-|d|^2/(2L))
\end{equation}
implying that at this order the divergence of the susceptibility in a finite dimensional lattice 
takes place in correspondence with the divergence in a Bethe lattice with connectivity $z=2D$.

\section{One spatial loop in the RS phase}

In a Bethe lattice in the thermodynamic limit, the density of loops of finite length vanishes, while spatial loops of finite length are common in finite dimensional lattices. In ref.~\cite{Mlayer} a new expansion around the Bethe lattice is performed.
As a result, the first correction to the bare propagator computed in the precedent sections comes from one spatial loop.
In this section, we will compute this contribution confirming that it is totally equivalent to the first correction computed in the usual field theoretical loop expansion, as stated in ref. \cite{Mlayer}.
Following the prescriptions of ref. \cite{Mlayer}, we construct a spatial loop structure $\mathcal{L}$, shown in Fig. \ref{Fig:loop},
formed by two paths of length $L_1+1$ and $L_2+1$ between the points $x$ and $y$ (the length of the internal paths are defined of length $L_1+1$ and $L_2+1$
because in this way results are more compact), plus two external legs of length $L_0$ and $L_3$ 
to the external spins $\sigma_0$ and $\sigma_d$.  The rest of the lattice is a Bethe lattice without loops in the thermodynamic limit.

We will compute the correction to the ``bare'' correlation functions computed in the previous section, that are those connected over the disorder.
Analogously to the definition of $T$, we define the vertex 
\small
\begin{align*}
V_{ab,cd,ef}(x)=&\overline{\langle\sigma_x^a\sigma_x^b\sigma_x^c\sigma_x^d\sigma_x^e\sigma_x^f\rangle}=\\
=&\begin{cases}
1 &\mbox{if  three pairs of indices are equal}\\                   
m_2 &\mbox{if  two pairs of indices are equal}\\                   
m_4 &\mbox{if  one pair of indices is equal}\\                   
m_6 &\mbox{if  the indices are all different}\\                   
\end{cases}
\end{align*}
\normalsize
As in the case of $T$, $a\neq b$, $c\neq d$, $e\neq f$.
We write the partition function in the presence of one loop $\mathcal{L}$ and we expand it for small $J$s analogously to eq. (\ref{eq:Zn_largez}).
From it, following the same reasoning of Sec. "The replica computation of line-correlations", we obtain the form of the correlation function when a structure $\mathcal{L}$ is present:
\begin{align}
\nonumber
\overline{\langle\sigma_0^a\sigma_0^b\sigma_d^c\sigma_d^d\rangle}^c_{\mathcal{L}}=&\overline{\langle\sigma_0^a\sigma_0^b\sigma_d^c\sigma_d^d\rangle}^c_{L_0,L_1+1,L_3}+\overline{\langle\sigma_0^a\sigma_0^b\sigma_d^c\sigma_d^d\rangle}^c_{L_0,L_2+1,L_3}+\\
&+\overline{\langle\sigma_0^a\sigma_0^b\sigma_d^c\sigma_d^d\rangle}^c_{lc}
\label{eq:oneloopcorr}
\end{align}
where the first and second terms turn out to be exactly the ``bare'' correlations computed as the two paths $L_0+(L_1+1)+L_3$ and $L_0+(L_2+1)+L_3$ were independent.
They have respectively $L_0+(L_1+1)+L_3$ and $L_0+(L_2+1)+L_3$ couplings, that are the minimal number of couplings to have a non zero correlation.
The last term has $L_0+(L_1+1)+(L_2+1)+L_3$ couplings and turns out to be
\begin{align}
 \nonumber
 \overline{\langle\sigma_0^a\sigma_0^b\sigma_d^c\sigma_d^d\rangle}^c_{lc}=&\((\frac{\beta^2}{2}\))^2\frac{1}{z^{L_0+(L_1+1)+(L_2+1)+L_3}}\times\\
&\times\sum_{q,r,s,t}\((T^{L_0}\))_{ab,qr}\left[\sum_{e,f,g,h,l,m,o,p}
V_{qr,ef,gh}\((T^{L_1}\))_{ef,lm}\((T^{L_2}\))_{gh,op}V_{st,lm,op}\right]\((T^{L_3}\))_{st,cd}.
\label{eq:loop_corr}
\end{align}
Thus in our case, from eq. (\ref{eq:oneloopcorr}) we see that $\overline{\langle\sigma_0^a\sigma_0^b\sigma_d^c\sigma_d^d\rangle}^c_{lc}$ is exactly the line-connected correlation function,
that gives the one-loop correction in the $\frac{1}{M}$ expansion around the Bethe solution.
The one loop contribution takes a very intuitive form.

To compute the explicit form for $\overline{\langle\sigma_0^a\sigma_0^b\sigma_d^c\sigma_d^d\rangle}^c_{lc}$,
we performed sums and products in eq. (\ref{eq:loop_corr}) using Mathematica
\footnote{The same results can be concluded from Eq.(62)
of Ref. \cite{Temesvari1} after a suitable correspondence between quantities like the
propagators and 3-point vertex.}.
At this point we can compute the one-loop contribution to the different correlation functions
as in Sec. ``The replica computation of line-correlations``. The whole expressions are reported in the Supplementary Material.
As for the bare term, the dominant terms are those with the highest power of $\lambda_R$, that are:

\begin{align}
 \overline{\((\langle\sigma_0\sigma_d\rangle_c\))^2}_{lc}\simeq(\overline{\langle\sigma_0\rangle^2\langle\sigma_d\rangle^2}^c)_{lc}\simeq32 \lambda_R^{L_0+L_1+L_2+L_3} [&1+44 m_2^2+101 m_4^2+m_4 (22-90 m_6)+\\
&-2 m_2 (7+67 m_4-30 m_6)-10 m_6+20 m_6^2] 
\label{eq:DominantTermLoop}
\end{align}

As explained in Sec. ``The dominant contribution in the correlation functions'', following ref. \cite{Mlayer}, 
we can now compute the correction to the susceptibility summing over all the length $L_0, L_1, L_2, L_3$, once we have multiplied 
by the number of non-backtracking walks. 

\section{Relation with previous RG studies}

In ref.~\cite{BR}, the authors perform standard RG calculations in $6-\epsilon$ dimensions for SG with a field.
They write the field-theoretic Hamiltonian in the vicinity of the critical line projected on the replicon eigenspace as
a function of the order parameter $q_{\alpha\beta}$ as:

\begin{align*}
H=&\frac{1}{4}r\sum q_{\alpha\beta}^2+\frac{1}{4}\sum (\nabla q_{\alpha\beta})^2+\\
&-\frac{1}{6} w_1\sum q_{\alpha\beta}q_{\beta\gamma}q_{\gamma\alpha}-\frac{1}{6} w_2\sum q^3_{\alpha\beta}
\end{align*}
with $r$ the reduced temperature, and $w_1$, $w_2$ being coupling constants. 
The correlation functions in the momentum space $k$ (projected on the replicon eigenspace) are proportional to $(r+k^2)^{-1}$.
Integrating over an infinitesimal shell $e^{-dl}<k<1$ in the momentum space, they obtain the recursion relation for $r$:

\begin{equation}
\frac{\text{d}r}{\text{d}l}=(2-\eta)r-\frac{K_d}{(1+r)^2}(4 w_1^2-16 w_1w_2+11w_2^2),
\label{Eq:BR}
\end{equation}
with $\eta=\frac{1}{3}K_d(4 w_1^2-16 w_1w_2+11w_2^2)$ and $K_d$ the usual geometrical factor, together with analogous recursion relations for $w_1$ and $w_2$.
The first term in Eq. (\ref{Eq:BR}) is the contribution connected to a renormalization of the critical temperature,
and that we can compute in our approach computing the correction given by a spatial tadpole structure. 
The second term is the one coming from the non-trivial loop. 

Following ref. \cite{w1w2Parisi},\cite{w1w2Sompolinsky}, in a Bethe lattice in finite and infinite connectivity, as well as in the fully-connected model, 
$w_1\propto1-3m_2+3m_4-m_6$ and $w_2\propto2m_2-4m_4+2m_6$.
Inserting these expressions in the loop term of eq. (\ref{Eq:BR}), we obtain:
\small
\begin{align*}
4 w_1-16 w_1w_2+11w_2^2\propto &4 - 56 m_2 + 176 m_2^2 + 88 m_4 +\\
&- 536 m_2 m_4 + 404 m_4^2 - 40 m_6 +\\ 
&+240 m_2 m_6 - 360 m_4 m_6 + 80 m_6^2
\end{align*}
that is exactly proportional to the coefficient of the dominant term
for the spatial-loop correction to the connected correlation function eq.~(\ref{eq:DominantTermLoop})\footnote{Please note that the replicon contribution, that is the one multiplied by the coefficient $b_4$, that corresponds to the
one found by Bray and Roberts, is the dominant one for the correlation functions. In our
approach, we obtain also the sub-dominant corrections coming from the other sectors, as in ref. \cite{Temesvari1,Temesvari2}.}.
Our approach permits to find the same results in a  clearer, simpler and more physically intuitive way.

\section{Conclusions and perspectives}

In ref.~\cite{Mlayer} a new expansion is introduced around the Bethe lattice. This expansion can be applied to all the models that can be defined on a Bethe lattice.
It is supposed to give the same results as standard perturbation theory for fully connected models when the physics of fully connected and finite connectivity MF models is the same.
We test the new expansion of ref. \cite{Mlayer} for the first time in the case of the spin glass model at finite temperature and with an external field in the limit of high connectivity.
We analytically find the same results of standard RG \cite{BR}, confirming the validity of the new $M$-layer topological expansion.

The new method has, however, the advantage that the equations are physically very intuitive: 
the bare correlations are just the correlation on a Bethe lattice (i.e.\ on a line), and
the first-order correction to the correlation function is just
the value of the correlation function computed on a spatial loop (finite loops are absent in the Bethe lattice and present in finite dimensional systems) once
the contributions of the two lines forming the loop, considered as independent graphs, are subtracted.
Even if for the case of the SG in a field in the limit $z\rightarrow \infty$ and finite temperature 
the results of the new expansion add nothing to what already known about the SGs in finite dimensions, there are cases in which things should be different.

It was already underlined how the Bethe lattice is more similar to finite dimensional systems \cite{CammarotaBiroliTarziaTarjus} with respect to the FC version 
for different disordered spin models. In the particular case of SG in a field, the critical line on the FC model tends to infinite field when the temperature goes to zero,
while in the Bethe lattice it ends at a finite field $h_c$ at $T=0$ \cite{w1w2Parisi}.
If the critical point for SGs in field in finite dimension is a zero temperature one, as supposed by some authors \cite{ParisiTemesvariT0dFinite}\cite{SG_MK}, it is crucial 
to perform an expansion around the Bethe solution, instead of around FC model, since the latter model is always in the SG phase at $T=0$.
This paper is a first step in this direction. The application of the topological expansion to the SG model in a field at $T=0$ and finite connectivity is at the moment under study. 
It could be done according to lines similar to what done in the case of the random field Ising model \cite{RFIM_Mlayer}.
\vspace{.5cm}

We thank Tommaso Rizzo for very useful discussions. We acknowledge funding from the European Research Council (ERC) under
the European Unions Horizon 2020 research and innovation programme
(grant agreement No [694925]).

\section{Calculation of $T^k$}\label{Appendix:T^K}

We are interested in correlation functions of the spin glass model in field, that are connected
with respect to the disorder. At this end it is convenient to compute 
\begin{equation}
 G_{a,b;c,d}(L)\equiv\overline{\langle\sigma_0^a\sigma_0^b\sigma_L^c\sigma_L^d\rangle}-\overline{\langle\sigma_0\rangle^2}\cdot \overline{\langle\sigma_L\rangle^2}\,,
\end{equation}
from which we can extract connected and disconnected (with respect to thermal average) correlation functions.

We define the matrix $T$ for $a\neq b$ and $c\neq d$ as:
\begin{align}
T_{ab,cd}(i)=&\frac{\beta^2}{2}\langle\sigma_i^a\sigma_i^b\sigma_i^c\sigma_i^d\rangle=\\
\nonumber
=&\frac{\beta^2}{2}\cdot 
\begin{cases} 
 1    &\text{ if }a=c\text{ , }b=d\text{ or if }a=d\text{ , }b=c \\
 m_2 &\text{ if }a=c\text{ or }b=d\text{ or }a=d\text{ or } b=c \\
 m_4 &\text{ if } a\neq b\neq c\neq d
\end{cases}
\end{align}
that can be written in the form:
\begin{align} 
 \nonumber
 T_{ab,cd}=&\frac{\beta^2}{2}\cdot\left[m_4+(m_2-m_4)(\delta_{ad}+\delta_{bc}+\delta_{bd}+\delta_{ac})+\right.\\
 &\left.+(1-2m_2+m_4)(\delta_{ac}\delta_{bd}+\delta_{ad}\delta_{cb})\right]
\end{align}
We need to compute the powers of $T_{ab,cd}$ in order to compute the correlation functions as explained in the main text:
\begin{equation}
G_{a,b;c,d}(L)=\frac{2}{\beta^2z^L}\left[\prod_{i=0}^LT(i)\right]_{ab,cd}.
\end{equation}

In order to diagonalize $T$, we look for eigenvalues and eigenvectors, of the form:
$\sum_{cd}T_{ab,cd}\,\psi_{cd}=\lambda\,\psi_{ab}$.
Because of the symmetry of the matrix $T$ under permutations of the replica indices, we know that there are three symmetry classs of eigenvectors 
(in an analogous way to what one does when looking to the stability of the Sherrington Kirkpatrick solution for the FC model \cite{dAT}):
The first one (longitudinal) is independent on the replica indexes, of the form: $$\psi_{ab}^{(L)}=C \quad \forall a,b \quad a\neq b.$$
Imposing $\sum_{cd}T_{ab,cd}C=\lambda_LC$ we find $\lambda_L(n)=\frac{\beta^2}{2}\cdot\left[(n-2)(n-3)m_4+4(n-2)m_2+2\right]$.
In the limit $n\rightarrow 0$: $$\lambda_L(0)=\beta^2(3m_4-4m_2+1).$$

The second $n-1$ eigenvectors (anomalous) depend just on one replica index and after simmetrization are of the form: 
$$\psi_{ab}^{(A)}=\frac{A_a+A_b}{2}  \quad a\neq b.$$
Imposing the orthogonality with respect to the first one we obtain $\sum_a A_a=0$.
Imposing $\sum_{cd}T_{ab,cd}\psi_{cd}^{(A)}=\lambda_A\psi_{ab}^{(A)}$ we find $\lambda_A(n)=\beta^2\left[(3-n)m_4+(n-4)m_2+1\right]$.

In the limit $n\rightarrow 0$: $$\lambda_A(0)=\beta^2(3m_4-4m_2+1).$$ There is a degeneration between $\lambda_L$ and $\lambda_A$
in the limit $n=0$.

The third $n(n-3)/2$ eigenvectors (replicon) depends on both the replica indexis and are of the form: $$\psi_{ab}^{(R)}=B_{ab} \quad a\neq b.$$
Imposing the orthogonality with respect to the second one we obtain $\sum_{b\neq a} B_{ab}=0$.
Imposing $\sum_{cd}T_{ab,cd}B_{cd}=\lambda_RB_{ab}$ we find $$\lambda_R=\beta^2(1-2m_2+m_4).$$

\subsection{Projectors on the eigenvectors}
We now want to express the matrix $T$ as a combination of projectors on the eigenvectors.
Thanks to the replica symmetry, we know that the important operators are those depending respectively on 0, 2 and 4 replica indexes:

\begin{align*}
R_{abcd}&=1\\
Q_{abcd}&=\frac{1}{4}\left[\delta_{ac}+\delta_{ad}+\delta_{bc}+\delta_{bd}\right]\\
P_{abcd}&=\frac{1}{2}\left[\delta_{ac}\cdot\delta_{bd}+\delta_{bc}\cdot\delta_{ad}\right]
\end{align*}

for which the following relations hold:
\begin{align*}
R\cdot R&=n(n-1)R\\
Q\cdot Q&=\frac{n-2}{2}Q+\frac{1}{2}R\\
Q\cdot R&=(n-1)R\\
P\cdot P&=P\\
P\cdot R&=R\\
P\cdot Q&=Q\\
\end{align*}

$R$ is proportional to the projector on the longitudinal eigenvector (that does not depend on replica indexes).
We define the normalized projector 
\begin{equation}
 \tilde{R}=\frac{1}{n(n-1)}R.
 \label{Eq:R}
\end{equation}
such that $\tilde{R}\cdot\tilde{R}=\tilde{R}$.
The projector on the space of anomalous eigenvectors
should be a combination of $R$ and $Q$ (because the anomalous eigenvector depends only on one replica index).
For this reason we define $\tilde{Q}=yQ-zR.$
Imposing the orthogonality with respect to $\tilde{R}$ and the normalization we find 
\begin{equation}
 \tilde{Q}=\frac{2}{n-2}\((Q-\frac{1}{n}R\)).
\label{Eq:Q} 
\end{equation}

The projector on the space of replicon eigenvectors
should be a combination of $P$, $R$ and $Q$ (because the replicon eigenvector depends on two replica indices).
Imposing the orthogonality with respect to $\tilde{R}$ and $\tilde{Q}$ and the normalization, we find the projector 

\begin{equation}
\tilde{P}=P-\frac{2}{n-2}Q+\frac{1}{(n-1)(n-2)}R.
\label{Eq:P}
\end{equation}

The inverse equations of eqs. (\ref{Eq:R},\ref{Eq:Q},\ref{Eq:P}) are:
\begin{align*}
R &= n(n-1)\widetilde{R}\\
Q &= \frac{n-2}{2} \widetilde{Q} + (n-1) \widetilde{R}\\
P &= \widetilde{P} + \widetilde{Q} + \widetilde{R}
\end{align*}

We can now rewrite $T$ as a combination of the projectors on the eigenvectors subspaces:

\begin{equation}
 T=\lambda_L\tilde{R}+\lambda_A\tilde{Q}+\lambda_R\tilde{P}.
\label{Eq:Tproj}
\end{equation}

Inserting the expressions for the eigenvalues and for the projectors in the previous expression, 
we check that we find again eq. (\ref{Eq:T}). 

\subsection{The limit $n\rightarrow 0$}
As found in eq. (\ref{eq:corr_d}), if we are interested in the correlations at distance $k$, we need to compute $T^k$. 
This is an easy task in the representation of eq. (\ref{Eq:Tproj}):
\begin{align}
 \nonumber T^k(n)=&\lambda_L^k\tilde{R}+\lambda_A^k\tilde{Q}+\lambda_R^k\tilde{P}=\\
=&R\left[\frac{1}{n}\((\frac{\lambda_L^k}{n-1}-\frac{2\lambda_A^k}{n-2}\))+\frac{\lambda_R^k}{(n-1)(n-2)}\right]
+Q\frac{2}{n-2}(\lambda_A^k-\lambda_R^k)+\lambda_R^kP.
\end{align}

We are interested in the limit $n\rightarrow0$, and in this limit $\lambda_L=\lambda_A$. So we have to treat carefully the term 
\begin{align*}
\lim_{n\rightarrow0}\frac{1}{n}\((\frac{\lambda_L^k}{n-1}-\frac{2\lambda_A^k}{n-2}\))=
&\frac{\text{d}}{\text{d}n}\left.\((\frac{\lambda_L^k}{n-1}-\frac{2\lambda_A^k}{n-2}\))\right\vert_{n=0}=\\
=&\frac{\beta^2}{2}k\lambda_{L/A}^{k-1}(0)\((3m_4-2m_2\))-\frac{\lambda_{L/A}^k(0)}{2}.
\end{align*}

At the end:

$$T^k(n=0)=\frac{\beta^2}{2}k\lambda_{L/A}^{k-1}\((3m_4-2m_2\))R+\lambda_{L/A}^k\((-\frac{R}{2}-Q\))+\lambda_R^k\((\frac{R}{2}+Q+P\)).$$

\section{The cavity computation for the bare correlations}

In this section we compute the bare correlation functions in a cavity approach, recovering the same results as the replica computation in the main text. 
The probability distribution on a linear chain of length $L$ is:
\be
P[\{\sigma_0,\ldots,\sigma_L\}] \propto \prod_{i=0}^L e^{\beta \((h_i^C+h_i^R\)) \sigma_i} \prod_{i=1}^L e^{\beta \sigma_{i-1} J_i \sigma_i}
\label{eq:Pcavity}
\ee
where we indicate with $h_i^C$ the cavity field, that is the sum of the contributions of all the links coming to the spin $i$ with the exception of the ones on the considered linear chain:
$$
h_i^C=\sum_{k\in \partial i,k\neq\{i-1,i+1\}} u_{k\rightarrow i}\,.
$$
The cavity bias $u_{k \to i}$ provides the marginal probability on spin $\sigma_i$ in the presence only of the neighbor $\sigma_k$, and satisfies the self-consistency equation:
\be
u_{i\rightarrow j}=\frac{1}{\beta}\text{atanh}\left[\text{tanh}\((\beta J_{ij}\))\text{tanh}\((\beta\sum_{k\in\partial i\backslash j}u_{k\rightarrow i}+\beta h^R_i\))\right]
\ee
A special care should be devoted to the spins $\sigma_0$ and $\sigma_L$ for which there is just one contribution coming 
from the linear chain.
The total field will instead be the sum of the contributions coming from all the neighbors:
$h_i=\sum_{k\in \partial i} u_{k\rightarrow i}$. 
The magnetization and the cavity magnetization are respectively:
$m_i=\text{tanh}(\beta h_i+\beta h_i^R)$, $m_i^C=\text{tanh}(\beta h_i^C+\beta h_i^R)$.

In the limit of high connectivity ($z\to\infty$) however $h_i^C\simeq h_i$, and in the high temperature phase we can write $m_i\simeq \beta ( h_i^C+h_i^R)$.
As before, we consider couplings: 
$\overline{J_{ij}}=0$, $\overline{J_{ij}^2}=1/z$. Higher orders are negligible in the $z\to\infty$ limit.
Thus we can write the probability distribution on a linear chain of length $L$, eq. (\ref{eq:Pcavity}), as
\be
P[\{\sigma_0,\ldots,\sigma_L\}] \propto \prod_{i=0}^L (1+m_i \sigma_i) \prod_{i=1}^L (1+\beta \sigma_{i-1} J_i \sigma_i)
\label{eq:cavity_P}
\ee
As for the replica computation, we must compute all the terms proportional to
$\prod_{i \in [1,L]} \beta^2 J_i^2,$
in the wanted correlation function, where again all the couplings should be present. 
This term can be computed explicitly as 
\begin{equation}
  \prod_{i=1}^LJ^2_i \prod_{i=1}^L\left (\frac{d^2}{dJ^2_{i}}\right)C\Big\rvert_{\{{J\}}=0}
  \label{eq:deriv_corr}
\end{equation}
where C is the expression for the desired correlation function.
We define $\chi_i \equiv 1-m_i^2$ and $\kappa_i \equiv 1-4m_i^2+3m_i^4 = (1-m_i^2)(1-3m_i^2)$ and we will use these definitions in the following.

Let us show how to perform the computation for the total-total correlation function. 
Combining eqs. (\ref{eq:cavity_P},\ref{eq:deriv_corr}) we obtain:
\scriptsize                    
\begin{align}
\label{eq1}
\nonumber
\langle \sigma_0 \sigma_L \rangle^2-\overline{\langle\sigma_0\rangle^2}\cdot \overline{\langle\sigma_L\rangle^2}=&\left[\frac{\sum_{\sigma_0,...,\sigma_L}\sigma_0 \sigma_L \prod_{i=0}^L (1+m_i \sigma_i) 
\prod_{i=1}^L (1+\beta \sigma_{i-1} J_i \sigma_i)}{\sum_{\sigma_0,...,\sigma_L}
\prod_{i=0}^L (1+m_i \sigma_i) \prod_{i=1}^L (1+\beta \sigma_{i-1} J_i \sigma_i)}\right]^2-\overline{\langle\sigma_0\rangle^2}\cdot \overline{\langle\sigma_L\rangle^2}\simeq\\
\nonumber
\simeq&\prod_{i=1}^LJ^2_{i}\prod_{i=1}^L\left(\frac{d^2}{dJ^2_{i}}\right)\left[\frac{\sum_{\sigma_0,...,\sigma_L}\sigma_0 \sigma_L \prod_{i=0}^L 
(1+m_i \sigma_i)\prod_{i=1}^L 
(1+\beta \sigma_{i-1} J_i \sigma_i)}{\sum_{\sigma_0,...,\sigma_L}\prod_{i=0}^L (1+m_i \sigma_i) \prod_{i=1}^L (1+\beta \sigma_{i-1} J_i \sigma_i)}\right]^2\Big\rvert_{\{{J\}}=0}\simeq\\
\simeq &\((\prod_{i=1}^L \beta^2 J_i^2\))\left[\sum_{i=0}^L \prod_{\substack{j=0\\j\neq i}}^L \kappa_j - L \prod_{i=0}^L \kappa_i + 
\prod_{i=0}^L \chi_i \left(A_0 + A_1 \sum_i m^2_i + A_2 \sum_{i<j} m^2_i m^2_j +\ldots \right)\right]
\end{align}
\normalsize
After some non-trivial algebra we find that the coefficients $A_k$ take values:
\be
A_k = (-1)^kA_2 \sum_{i=1}^{k-1} i 3^{i-1} = (-1)^k\left[(2k-3)3^k+3\right]
\ee
Thanks to this analytic expression, we can obtain the following expression for the total correlation once we have 
taken the average over the distribution of the $m_i$:
\be
 \overline{\langle \sigma_0 \sigma_L \rangle^2}^c\simeq \((\frac{\beta^2}{z}\))^L
\left[3(1-2m_2+m_4)^{L+1}+\Big((-2+6m_2-3m_4)+L(3m_4-2m_2)\Big) (1-4m_2+3m_4)^L\right]
\label{Eq:corr_cav}
\ee
where $m_2=\overline{m^2}$ and $m_4=\overline{m^4}$ are exactly the same ones appearing in the replica calculation of the precedent section.
Now it is simple to show that eq. (\ref{Eq:corr_cav}) is exactly the same as obtained with the replica method.




In the same way we can compute the
connected-disconnected correlation function:
\begin{align}
 \nonumber
\langle \sigma_0 \sigma_L \rangle_c\langle \sigma_0\rangle\langle \sigma_L \rangle-\overline{\langle\sigma_0\rangle^2}\cdot \overline{\langle\sigma_L\rangle^2}=&
\beta^{2L}\prod_{i=1}^LJ^2_{i,i-1}\prod_{i=0}^L \chi_i\left[a_1\sum_{i=0}^L m_i^2+\right.\\
&\left.+a_2\sum_{i<j=0}^L m_i^2m_j^2+a_3\sum_{i<j<k=0}^L m_i^2m_j^2m_k^2+...\right]
\label{Eq:discConRS}
\end{align}

with coefficients $a_i=(-1)^{i+1}\sum_{x=0}^{i-1}3^x=(-1)^{i+1}(-\frac{1}{2}+\frac{3^i}{2})$. 
Once we have taken the average over the distribution of the $m_i$, we obtain the final expression:

\begin{align*}
\overline{\langle \sigma_0 \sigma_L \rangle_c\langle \sigma_0\rangle\langle \sigma_L \rangle}^c=\((\frac{\beta^2}{z}\))^L\frac{1}{2}\left[(1-2 m_2+m_4)^{L+1}-(1-4 m_2+3m_4)^{L+1}\right].
\end{align*}
We can derive the same expression in the replica approach.

The expression for the connected-connected correlation function, computed in the same way, is:
\begin{align}
 \((\langle \sigma_0 \sigma_L \rangle_c\))^2=\beta^{2L}\prod_{i=1}^LJ^2_{i,i-1} \prod_{i=0}^{L} \chi_i^2.
\label{Eq:ConConRS}
\end{align}
Averaging over the magnetizations we find again the replica results.

\section{One loop correction for the correlation functions}\label{Appendix:Coefficients}

In the main text, we explained how to compute the one-loop contribution to the different correlation functions.
The whole expressions depending on all the eigenvalues of $T$ are:
\begin{itemize}
 \item For the connected-connected correlation function:
\begin{align}
 \nonumber
\overline{\((\langle\sigma_0\sigma_d\rangle_c\))^2}_{lc}=&\lambda_R^{L_0+L_3}\left[\left(L_2 \lambda_{L/A}^{L_2-1} \lambda_R^{L_1}+L_1 \lambda_{L/A}^{L_1-1} \lambda_R^{L_2}\right)b_1 \beta^2
+\lambda_{L/A}^{L_1+L_2}b_2 \right.\\
&\left.+(\lambda_{L/A}^{L_2} \lambda_R^{L_1}+\lambda_{L/A}^{L_1} \lambda_R^{L_2})b_3+\lambda_R^{L_1+L_2} b_4\right]
\label{Eq:CCloop}
\end{align}

\item For the disconnected-disconnected correlation function:

\begin{align}
\nonumber
(\overline{\langle\sigma_0\rangle^2\langle\sigma_d\rangle^2}^c)_{lc}=&c_1\beta^4  \left[L_0 (L_1 + L_2) + L_2 L_3 + L_1 (L_2 + L_3)\right] \lambda_{L/A}^{L_1 + L_2 + L_3 + L_0 - 2} + \\ 
\nonumber 
&+\beta^2 \lambda_{L/A}^{L_1 + L_2 + L_3 + L_0 - 1}\((c_2(L_0+L_3)+c_3(L_1 + L_2)\))\\
\nonumber
 &+c_4 \lambda_{L/A}^{L_1 + L_2 + L_3 + L_0} + 
 c_5  \lambda_R^{L_1 + L_2 + L_3 + L_0} +\\ 
\nonumber
 &+c_6\beta^2 (L_1 \lambda_{L/A}^{L_1 - 1} \lambda_R^{L_0 + L_2 + L_3} + 
    L_2 \lambda_{L/A}^{L_2 - 1} \lambda_R^{L_1 + L_3 + L_0}) + \\
\nonumber
    &+c_7 (\lambda_{L/A}^{L_1} \lambda_R^{L_0 + L_2 + L_3} + \lambda_{L/A}^{
     L_2} \lambda_R^{L_1 + L_3 + L_0}) + \\
\nonumber
     &+c_8\beta^2 (L_0+L_3) \lambda_{L/A}^{L_0 + L_3 - 1}  \lambda_R^{L_1 + L_2} + \\
\nonumber
 &+c_9 (\lambda_{L/A}^{L_2 + L_3 + L_0} \lambda_R^{L_1} + \lambda_{L/A}^{
     L_1 + L_3 + L_0} \lambda_R^{L_2}) +\\ 
 &+c_{10} \lambda_{L/A}^{L_0 + L_3} \lambda_R^{L_1 + L_2} + 
 c_{11} \lambda_{L/A}^{L_1 + L_2} \lambda_R^{L_0 + L_3}.
 \label{Eq:DDloop}
\end{align}

\end{itemize}

We compute the numerical value of the coefficients of the different terms, depending on $m_2$, $m_4$, $m_6$ in the Gaussian approximation.
In fact for $z=\infty$, the $a$-th moment of the magnetization $m_a$ can be computed solving the equation:
\be
m_a=\frac{1}{\sqrt{2\pi}} \int_{-\infty}^{\infty}e^{-t^2/2}\tanh^a\((\beta t \sqrt{Q}+\beta h\))\text{d}t
\ee
imposing that $Q=m_2$, with $h$ the external field. 

In this limit, all the coefficients are different from 0 when the field is present. In the absence of field the coefficients $b_1, c_1, c_2, c_3, c_6, c_8$ are null. 
Please note that they are the coefficients that multiply factors $L$. However in this limit, the tree eigenvalues $\lambda_R$, $\lambda_{L/A}$ are degenerate and
things should be computed in a more careful way 
\footnote{In the limit of small magnetic field, also the standard field theory has degenerate eigenvalues and the Bray-Roberts solution (see following Section) is no more valid.}.

We write in the following the explicit expressions for the coefficients in eqs. (\ref{Eq:CCloop},\ref{Eq:DDloop}):

\begin{align*}
b_1=&-32 (2 m_2-3 m_4) (1-7 m_2+11 m_4-5 m_6)^2 \\
b_2=&64 (1-7 m_2+11 m_4-m_6)^2 \\
b_3=&-80 (1+35 m_2^2+77 m_4^2+m_4 (18-68 m_6)-2 m_2 (6+52 m_4-23m_6)-8 m_6+15 m_6^2)\\
b_4=&32 \left(1+44 m_2^2+101 m_4^2+m_4 (22-90 m_6)-2 m_2 (7+67 m_4-30 m_6)-10 m_6+20 m_6^2\right) 
\end{align*}

\begin{align*}
c_1=&64 (2 m_2-3 m_4)^2 (2-17m_2+30 m_4-15 m_6)^2\\
c_2=&32 (2 m_2-3 m_4) (-2+17 m_2-30 m_4+15 m_6)(-2+8 m_2-6 m_4)\\
c_3 =& -96 (2 m_2 - 3 m_4) (-2 + 17 m_2 - 30 m_4 + 15 m_6) (3 m_2 - 8 m_4 + 5 m_6)\\
c_4=&-8 (13 + 972 m_2^2 + 2493 m_4^2 - 2 m_2 (116 + 1578 m_4 - 705 m_6) +\\ 
   &-180 m_6 + 450 m_6^2 - 30 m_4 (-13 + 72 m_6))\\
c_5=&32 (1 + 44 m_2^2 + 101 m_4^2 + m_4 (22 - 90 m_6) - 
   2 m_2 (7 + 67 m_4 - 30 m_6) - 10 m_6 + 20 m_6^2)\\
c_6=&-32 (-1 + 7 m_2 - 11 m_4 + 5 m_6) (2 m_2 - 3 m_4) (-1 + 7 m_2 - 11 m_4 + 
   5 m_6)\\
c_7=&-40 (-1 + 7 m_2 - 11 m_4 + 5 m_6) (-1 + 5 m_2 - 7 m_4 + 3 m_6)\\
c_8=&-96 (-1 + 7 m_2 - 11 m_4 + 5 m_6) (2 m_2 - 3 m_4) (-1 + 7 m_2 - 
   11 m_4 + 5 m_6)\\
c_9=&96 (-1 + 7 m_2 - 11 m_4 + 5 m_6) (-1 + 7 m_2 - 11 m_4 + 5 m_6)\\
c_{10}=&-24 (-1 + 7 m_2 - 11 m_4 + 5 m_6) (-3 + 11 m_2 - 13 m_4 + 5 m_6)\\
c_{11}=&32 \((1 - 7 m_2 + 11 m_4 - 5 m_6\))^2 
\end{align*}

\end{document}